# Phase separation of a supersaturated nanocrystalline Cu-Co alloy and its influence on thermal stability

A. Bachmaier[1*], M. Pfaff[2], M. Stolpe[3], H. Aboulfadl[4], C. Motz[1]

[1] Chair of Materials Science and Methods, Saarland University, Saarbrücken, Germany
[2] INM-Leibniz Institute for New Materials, Saarbrücken, Germany
[3] Chair of Metallic Materials, Saarland University, Saarbrücken, Germany
[4] Chair of Functional Materials, Saarland University, Saarbrücken, Germany
*Corresponding author: a.bachmaier@matsci.uni-sb.de

**Abstract**
The thermal decomposition behavior, the microstructural evolution and its influence on the mechanical properties of a supersaturated Cu-Co solid solution with ~100 nm average grain size prepared by severe plastic deformation is investigated under non-isothermal and isothermal annealing conditions. Pure fine-grained Cu and Co exhibit substantial grain growth upon annealing, whereas the Cu-Co alloy is thermally stable at the same annealing temperatures. The annealed microstructures are studied by independent characterization methods, including scanning electron microscopy, electron energy loss spectroscopy and atom probe tomography. The phase separation process in the Cu-Co alloy proceeds by the same mechanism, but on different length scales: a fine scaled spinodal-type decomposition is observed in the grain interior, simultaneously Co and Cu regions with a larger scale are formed near the grain boundary regions. Subsequent grain growth at higher annealing temperatures results in a microstructure consisting of the pure equilibrium phases. Such mechanisms can be used to tailor nano structures to optimize certain properties.

## 1. Introduction

A long-standing interest exists in metastable solid solutions due to their potential to achieve outstanding mechanical and physical properties [1-3]. In recent times new methods have been developed and improved to obtain supersaturated solid solutions. For example, severe plastic deformation (SPD) by high-pressure torsion (HPT) deformation has been used to produce bulk, nanocrystalline (nc) solid solutions in a variety of binary Cu-based alloy systems with a positive heat of mixing (for instance: Cu-Fe, Cu-Cr, Cu-Co, Cu-W [4-13]) by extending the solid solubility level during deformation. As expected, these metastable solid solutions are not in the equilibrium state after HPT processing and during subsequent annealing treatments, decomposition of the metastable solid solutions is reported [4,9,10,12]. Tailoring material properties for potential technological applications requires an understanding of the thermal stability, the structural changes (phase decomposition, nucleation and growth) during annealing after HPT processing and the atomic-scale processes underlying elemental decomposition.

Common approaches to improve the thermal stability of nc materials with a binary composition include pinning of grain boundaries by solutes or lowering their grain boundary energy by segregation of these solute atoms to the grain boundaries (i.e. kinetic and thermodynamic approach, respectively) [14-22]. If grain boundary segregation upon annealing is the dominant process, it might result in an idealized nanostructure, where a bulk nc grain is surrounded by a phase consisting solely of solute atoms. Grain growth would be theoretically not possible in such structures with a positive heat of mixing since the latter prevents re-dissolution of solute atoms in the matrix [23]. In addition, grain boundary migration kinetics are lowered by the segregated solutes. Such a configuration of two, immiscible phases, has already been reported to stabilize the material effectively against grain growth in supersaturated nc Nb-Cu [24]. In [25], nanostructure stability maps based on a thermodynamic model are introduced, in which the segregation strength relative to mixing enthalpy of a binary system determines the propensity for segregation stabilization [26, 27].

The Cu-Co system is a relatively simple peritectic alloy system with a positive heat of mixing and a wide metastable region. Below 400°C the solubility of Co in Cu (and vice versa) is negligible [28]. Homogeneous Cu-Co alloys with supersaturations far from equilibrium can be produced by ball milling [29, 30], sputtering [31], melt-spinning [32] and rapid quenching [33]. Since the Cu-Co system has a large positive heat of mixing, there is a strong tendency for decomposition of the supersaturated solid solution at elevated temperatures, which has been widely studied by means of many techniques [34-37]. In addition, it has been shown that in $Cu_{90}Co_{10}$ alloys, a spinodal type decomposition of the formed supersaturated solid solutions occurs after annealing at elevated temperatures [32, 33].

Regarding physical properties, the Cu-Co system is also an excellent candidate for the giant magnetoresistance effect [38], where the size and morphology of the embedded Co phase has a huge influence on the electric resistivity under an applied magnetic field [39-41]. Berkowitz et al. [31] found that the giant magnetoresistance effect can be observed in Cu-Co alloys with a granular structure composed of nanometer sized Co precipitates dispersed in the nonmagnetic Cu matrix. Later on, it was proposed that the giant magnetoresistance effect in Cu-Co alloys originate from modulated microstructures evolving from a homogeneous spinodal decomposition inside the grains in melt-spun ribbons [32].

Recently, we reported that the formation of supersaturated solid solutions in the Cu-Co alloy system is also feasible by HPT deformation [11]. In the studied alloy, large supersaturations up to 26 at.% Co in fcc Cu with an average grain size of 101±20 nm are obtained in the as-deformed state (for further details see Ref. [11]). In comparison with other processing techniques [29-33], the HPT process has certain advantages as the alloying process proceeds in the solid state. Motivated by the promising properties such as high strength, high thermal stability and the giant magnetoresistance effect, the Cu-Co system is chosen as model material to elucidate thermal stability, thermal decomposition and microstructural evolution during non-isothermal and isothermal annealing conditions (between ~25% and ~75% of $T_m$) of metastable solid solutions after HPT processing in detail in this study. Compared to aforementioned studies on decomposition of supersaturated solid solutions in Cu-Co alloys [33-37], the subjects of the present study are to investigate the effect of higher Co content (up to 26 at.%), the influence of the processing technique (SPD) which induces a different defect structure and the nanometer sized grains in the as-processed state on decomposition and thermal stability. Decomposition mechanism and kinetics were studied using differential scanning calorimetry (DSC) and differential thermal analysis (DTA). Changes in microstructure during annealing are evaluated using scanning electron microscopy (SEM), atom probe tomography (APT) and transmission electron microscopy (TEM). APT was used here since it offers substantial resolution in 3D at the atomic-scale for determining local elemental decomposition along interfaces [42,43]. Novel results obtained from these experiments are compared to the thermal stability of single-phase HPT deformed pure Cu and Co reference materials and will be discussed in terms of common approaches to enhance the stability of nc materials. It is shown that phase separation proceeds similar to spinodal decomposition on different length scales at intermediate temperatures (~40% $T_m$). The evolving microstructure –fine-scaled compositional modulated grains together with pure Co and Cu particles - is stable even for very long annealing times of several days. In contrast, the metastable solid solution quickly transforms back to the pure elemental equilibrium phases at higher annealing temperatures, which is accompanied by pronounced grain growth.

## 2. Experimental

A binary Cu-based alloy containing 26 at.% Co and pure Co and Cu as reference materials were prepared by RHP-Technology (Seibersdorf, Austria) from Co powders (purity 99.8%) and Cu powders (purity 99.9%) and precompacted into samples with cylindrical shape. Disks of the alloy and pure reference materials with a thickness of 0.6 mm and a diameter of 8 mm were subjected to SPD by HPT at room temperature (pressure of 5 GPa, 25 rotations with a rotation speed of 0.2 rpm). The shear strain produced by torsion straining can be given as von Mises equivalent shear strain $\varepsilon_{eq}$ and is calculated according to [44]

$$\varepsilon_{eq} = \frac{2*\pi*r*N}{\sqrt{3}*t},  \qquad (1)$$

with N being the number of applied rotations during HPT deformation, r being the distance from the center of the sample, t being the sample thickness. All samples for each of the following investigations were examined at a distance ≥ 1.5 mm from the sample centre. For this distance, the equivalent shear strain value $\varepsilon_{eq}$ is ≥ 210. This strain value has been typically used to obtain the steady state region of the microstructural refinement for the alloy and the pure reference materials in earlier studies, in which application of further HPT deformation does not induce further grain refinement [11].

To determine the heat release of the as-deformed samples, DSC and DTA measurements were performed using a Perkin Elmer DSC 8000 and a Netsch STA 446/C/6/MFG/G Jupiter device, respectively. For all DSC and DTA measurements, a subsequent re-run served as the reference measurement and baseline for the analysis. For the determination of the activation energy of grain growth $Q$ for the pure reference materials with the method of Kissinger, DSC measurements with different linear heating rates ($\Phi$) of 10, 20, 60 and 120 °C/min in a temperature range from 25°C to 720°C were performed on all deformed samples. By evaluating the shift of the peak temperature ($T$) for different heating rates $\Phi$ and plotting

$$\ln\left(\frac{\Phi}{T^2}\right) = -\left(\frac{Q}{RT}\right) + const, \qquad (2)$$

with R being the gas constant, the activation energies $Q$ can be determined.

For SEM investigations of the microstructure, samples were prepared from the same part of the HPT disks as the DTA and DSC samples and subsequently annealed up to predefined temperatures (250°C, 720°C, 1000°C) at a heating rate of 20 °C/min, followed by rapid cooling to ambient temperatures. Additionally, the processed samples were subjected to isothermal heat treatments at 150°C, 300°C, 400°C and 600°C for 1 h, 4 h, 7 h, 27 h and 100 h after HPT deformation to investigate the long-term thermal stability of the Cu-Co alloy and the pure reference materials. The Vickers microhardness HV0.5 (as-deformed and isothermally annealed samples) was measured using a 500 g load and is the average value of 15 measurements. Microstructural characterization in as-deformed and annealed conditions was carried out by SEM using a Zeiss SIGMA™-VP field emission scanning electron microscope device, which is equipped with a backscattering electron (BSE) detector and an energy dispersive spectroscopy (EDS) detector. The microanalysis data was evaluated using the AZtec software (Oxford Instruments). TEM investigations were performed on a JEOL JEM 2011 and a cold field emission gun TEM/STEM (JEOL JEM-ARM 200F) instrument. For microstructural observation in bright-field (BF) and dark-field (DF) modes, to record selected-area electron diffraction (SAD) patterns and high-resolution Electron Energy Loss Spectroscopy (EELS) maps, both instruments were operated at an accelerating voltage of 200 kV. The JEOL JEM-ARM 200F instrument is equipped with a STEM Cs corrector (CESCOR; CEOS GmbH Heidelberg). EELS was performed using a post-column EELS spectrometer (GIF QuantumER™ from Gatan). For STEM, the annular dark-field (ADF) detector was used with a camera length of 8 cm, resulting in a collection angle range of 70-280 mrad. The EELS measurements were carried out with a beam current of 530 pA, a collection semi angle of 10.4 mrad, a spectrometer entrance aperture of 2.5 mm and a dispersion of 0.25 eV/ch. A drift correction was applied to enhance the image quality. The TEM sample preparation includes the following steps: Disks of deformed and selected annealed samples were cut, mechanically thinned and polished to a thickness of about 100-120 μm. Subsequently, mechanical dimpling until the thinnest part reaches a thickness of about 10 μm was conducted. In a final step, the TEM samples were ion-milled with Ar ions at 4-5 kV under an incidence angle of 5-7° using a Gatan Precision Ion Polishing System until perforation was obtained.

APT measurements were carried out using a LEAP™ 3000X HR CAMECA™ system for the as-deformed state sample and for the 100h at 400°C annealed sample, respectively. Site specific sample preparations were done in a dual-beam focused ion beam/SEM workstation implementing the in-situ liftout technique [45]. A 200 nm thick electron beam Pt-cap layer was first deposited by physical vapor deposition for reducing Ga implantation during specimen preparation. Subsequently, 2 keV ion energy was used for final steps during specimen shaping to minimize Ga induced damage. The specimens were measured in laser pulsing mode (532 nm wavelength) with a repetition rate of 160 kHz, a base temperature of 60-65 K and laser pulse energy of 0.5 nJ. The data were reconstructed using the standard algorithm developed by Bas et al. [46] and analyzed with the software CAMECA$^{TM}$ IVAS 3.6.8.

## 3. Results

During HPT deformation of the binary Cu-Co alloy and the pure Co and Cu reference materials, significant grain refinement and a considerable increase in hardness is observed. The microhardness of pure Cu, pure Co and the Cu-Co alloy in the as-deformed state is 207±4 HV0.5, 430±37 HV0.5 and 281±4 HV0.5, respectively. The grain size, which can be achieved in the steady state regime of HPT deformation, is truly nc for pure Co and ultra-fine grained (ufg) for pure Cu. Large supersaturations up to 26 at.% Co with an average grain size of 101±20 nm are obtained in the Cu-Co alloy in the as-deformed state [11].

*3.1 Non-isothermal annealing*

In the beginning, thermal analysis was performed with DSC and DTA measurements to determine decomposition temperature and annealing kinetics of the HPT deformed Cu-Co alloy and the pure Co and Cu reference materials. The as-deformed Cu-Co alloy samples show broad and overlapping exothermic peaks in the DSC scans. An example for a typical DSC scan at a heating rate of 20°C/min is shown in Fig. 1a. Overlapping peaks are common features observed in DSC scans during the decomposition of highly supersaturated solid solutions [29,47-49]. It is believed that the small heat release in the first part of annealing in the DSC scans is due to recovery processes. To clarify this assumption, a DSC scan on a sample that was heated to 250°C (below the onset of the main exothermic peak), held for 1 min at this temperature and subsequently cooled down to room temperature, was performed. The pre-annealed sample exhibit a smaller shoulder-shaped signal prior to the distinct second exothermic peak (Fig. 1a). The DSC results further indicate that the thermal processes of the Cu-Co alloy are not completed at the maximum temperature (720°C) which can be applied during the DSC measurements. Therefore, additional DTA measurements were carried out with a linear heating rate of 20 °C/min in a temperature range from 25°C to 1000°C (Fig. 1b). In the DTA scan, a second broad exothermic peak extending from ~700°C to the maximum temperature of the DTA scan (1000°C) is visible. To correlate thermal behavior with microstructural evolution, Cu-Co alloy samples were characterized by SEM after annealing from room temperature to 250°C, 720°C and 1000°C and quenching to room temperature as indicated by the red arrows in Fig. 1a-b. No change regarding the grains size between as-deformed and annealed microstructure is observable after annealing to 250°C (Fig. 1c). After annealing to 720°C, only minor grain growth seems to take place, although a substantial amount of heat has been released as visible from the DSC and DTA data for temperatures below 720°C. It becomes obvious from the microstructure of the sample annealed to 1000°C that the broad annealing stage visible in the DTA scan is accompanied by significant grain growth. Due to the micrometer sized grains, elemental mapping by EDS analysis can be performed on the sample annealed to 1000°C. In the small inset, the elemental distribution of Co and Cu is illustrated with a color overlay showing the variations in the X-ray spectrum. Regions appearing red in the micrograph represent the Cu phase, regions appearing green consist of the Co phase. It is apparent that besides pronounced grain growth, decomposition of the supersaturated solid solution during annealing occurred. In combination with the microstructural observations, it can be further concluded that the small part of the heat release at low temperatures can be reasonably attributed to the recovery to the HPT induced crystal lattice defects and relaxation of grain boundaries. These features are typically observed during annealing of HPT deformed materials [50].

For the pure Co and Cu reference materials, a weak shoulder-shaped signal prior to one distinct exothermic peak is recorded in the DSC scans. A list of peak temperatures $T$ for the main exothermic peak at various heating rates for pure Cu and Co can be found in Table 1. To correlate the heat releases observed during DSC scanning with modifications of the microstructure, characterization by SEM was further performed. Pure Co and Cu samples were annealed from room temperature to 250°C (a temperatures below the onset of the exothermic peaks) and to 720°C (a temperature above the exothermic peaks), and subsequently quenched to room temperature. Fig. 2 shows the microstructure of these samples in comparison to the as-deformed microstructures. No significant difference regarding the grain size between as-deformed (Fig. 2a,d) and annealed microstructure to a maximum temperature of 250°C is visible for both materials (Fig. 2b,e). In contrast, considerable grain growth takes place after annealing to 720°C (Fig. 2c,f). Analyzing the shift of the peak temperatures with heating rate according to Kissinger provides insights into the kinetics of grain growth observed for the pure reference materials. The analysis results in a very good linear relationship for all sets of data (see Supplementary Fig. S1), yielding the activation energies as listed in Table 1. For pure Cu, an activation energy $Q$ of 0.96±0.05 eV is determined, which is in excellent agreement with previous studies on HPT deformed Cu [51-53]. Analysis for pure Co yield a higher activation energy of $Q$=1.83±0.12 eV.

*3.2. Evolution of microstructure and microhardness during isothermic annealing*

The influence of long-term isothermal annealing (annealing time up to 100h) of the Cu-Co alloy on the mechanical properties and microstructural evolution is studied by microhardness measurements and first SEM and TEM analysis. Annealing temperatures are chosen according to the results of the DSC/DTA scans (below, at the onset, near the peak temperature and at the end of the distinct exothermic peak, respectively) and are indicated by the gray arrows in Fig. 1a-b. For comparison, the pure reference materials are isothermally annealed for 1h at 300°C, 400°C and 600°C as well.

In Fig. 3a, the evolution of the microhardness as a function of the annealing time and temperature is shown for the Cu-Co alloy sample. At low annealing temperatures (150°C-400°C), no significant change in the microhardness can be detected even after annealing for very long times (up to 100 h). Upon annealing at the highest annealing temperature (600°C), the microhardness is continuously decreasing with increasing annealing time. After annealing for 7 h at 600°C, a constant microhardness is reached, which is not further decreasing even for the longest annealing time of 100 h. The microhardness as a function of the annealing temperature of the pure reference materials is plotted in Fig. 3b. The microhardness of the pure Cu sample decreases even at the lowest annealing temperature of 300°C indicating grain growth. At higher annealing temperatures, the microhardness continuously decreases. The microhardness of the pure Co sample stays nearly constant for annealing at 300°C. At higher annealing temperatures, the microhardness drops indicating coarsening of the nanostructure.

Fig. 4 shows a sequence of BSE micrographs for different annealing temperatures and times in comparison to the as-deformed microstructure of the Cu-Co alloy sample. Each of these micrographs, except the one after annealing for 100h at 600°C, is of the same magnification and illustrates the microstructural evolution at different durations of isothermal annealing. The contrast in the BSE images depends on both, the grain orientation and the order number, hence there is no simple discrimination between Cu and Co possible. After annealing at 150°C for different times, no significant change or coarsening of the microstructure takes place compared to the as-deformed microstructure. A similar microstructure is depicted in the BSE micrographs of the specimens annealed at 300°C. However, the minor elongation of the grains in the HPT shearing direction observable at lower annealing temperature and in the as-deformed state is reduced. After annealing at 400°C for 1h, minor structural coarsening of some grains is visible in the BSE micrographs. These large grains exhibit a rather constant gray shade. Distinct microstructural features can be additionally found in the micrographs. In some of the grains, substructures with a very fine scaled variation in their shade are visible. Furthermore, small structural elements with a distinct different grayscale value are observable, which are located inside grains and at grain boundaries. These features become more clearly visible after annealing for 7h and 100h at 400°C. After annealing at 600°C for 1h, pronounced grain growth occurs. Grains with constant gray values are still visible in the BSE micrograph. Again, substructures with small-scaled variations in their gray shade can be observed. Compared to the microstructure after annealing at 400°C, these regions occur to a larger extent. Annealing for longer times at the highest annealing temperature lead to further distinct grain growth. In the BSE micrographs of the sample annealed for 100h at 600°C, the grains exhibit rather constant gray values, their size is about 400 – 500 nm and no modulated substructures are visible any more. Small structural elements with distinct different grayscale value which are located inside grains and at grain boundaries are furthermore clearly observable.

In Fig. 5 and 6, TEM micrographs illustrate the microstructure in the as-deformed and annealed state of the Cu-Co alloy after annealing for the longest annealing time of 100h at 300°C, 400°C and 600°C. Additionally, phase identification was performed by electron diffraction. Fig. 5 shows the BF micrograph (a), SAD pattern (b) and DF micrograph (c) of the as-deformed Cu-Co alloy sample. An ufg, non-distinct and difficult to resolve microstructure with a huge amount of defects is visible, which makes a clear identification of the grains and grain boundaries difficult. The ufg structure is also reflected in the SAD pattern. The spots form almost continuous diffraction rings, which suggest the existence of a large quantity of high-angle boundaries. In the SAD pattern, *fcc* Cu and *fcc* Co Debye-Scherrer rings of {111}, {200}, {220}, {311} and {222} planes are indicated. No *hcp* Co Debye-Scherrer rings are visible. The DF micrograph was obtained by selecting parts of the first Debye-Scherrer rings (most likely *fcc* Cu {111}) as indicated in the SAD pattern. Compared to the as-deformed microstructure, the grain size is not significantly changed after annealing at 300°C (Fig. 5d,f). In the BF and DF micrographs of the sample annealed at 400°C, minor structural coarsening is observed (Fig. 5 g,i). Both DF micrographs were recorded by selecting parts of the first Debye-Scherrer rings (*fcc* Cu {111} or *fcc* Co {111} or both) as indicated. Due to the small difference in the lattice parameters of *fcc* Cu and *fcc* Co, a discrimination of these reflection rings in the micrograph is not possible. In the SAD pattern, which is recorded after annealing at 300°C, only *fcc* Cu and *fcc* Co Debye-Scherrer rings of {111}, {200}, {220}, {311} and {222} planes are observed (Fig. 5e), whereas the pattern of the sample annealed at 400°C (Fig. 5h) reveals two additional weak *hcp* Co {102} diffraction spots (indicated by white circles). Isothermal annealing at an annealing temperature of 600 °C for 100h results in a microstructure, in which well-defined large grains coexist with some smaller grains (Fig. 6). Furthermore, annealing twins have formed. It can also be seen from Fig. 6a, that the dislocation density is very low in most of the large grains. The SAD pattern of the sample shows three sets of spotty diffraction rings including the *fcc* Cu, *fcc* Co and *hcp* Co phase (Fig. 6b). The DF micrographs were recorded by selecting a *fcc* Cu {200} or *fcc* Co {200} diffraction spot (Fig. 6c) and a *fcc* Co {220} or *hcp* Co {110} diffraction spot (Fig. 6d), respectively. In Fig. 6c, a large, single grain is visible in the DF micrograph. In contrast, very small structural features, which are located between two grains, at a triple point, and presumably inside a larger grain can be seen in the DF micrograph displayed in Fig. 6d.

From the TEM data, it can be assumed that the single phase *fcc* structure, which is formed in the as-deformed state, is maintained even after annealing for 100h at 300°C. At higher annealing temperatures, a microstructure consisting of three phases (*fcc* Cu, *fcc* Co and *hcp* Co) forms. Additionally, elemental mapping in the SEM by EDS for the Cu-Co alloy sample annealed at 600°C for 100h was performed (Fig. 7). The elemental distribution of Co and Cu is illustrated with a color overlay, in which areas appearing red in the micrograph represent Cu and areas appearing green consist mainly of Co. Visible Co regions can be identified by EDS which correlate to grains with darker contrast in the corresponding BSE micrograph. From the EDS mapping it is evident that phase separation occurred after long-term isothermal annealing at 600°C.

*3.3 Analytical transmission electron microscopy and atom probe tomography*

From the SEM and TEM investigations, there is not enough information to get a clear picture of the complex microstructure, which develops during isothermal annealing at intermediate annealing temperatures (400°C) or short annealing times (≤7h) at 600°C. Hence, the microstructure of the Cu-Co alloy is investigated in detail by analytical TEM and APT. An annealing temperature of 400°C and a time of 100h was chosen with respect to the high hardness of the Cu-Co alloy at this annealing state (hardness of 295±11 HV0.5), which is even slightly higher as the as-deformed state (281±4 HV0.5) indicating a pronounced decomposition without distinct grain growth.

Using EELS mapping in STEM mode, a large volume (compared to APT) of the sample can be analyzed which includes information about the chemical composition of the microstructure. In Fig. 8a-b, EELS Cu and Co maps with an area over 1500 x 1200 nm² are displayed. The microstructure looks similar to the ones observed in the BSE micrographs (Fig. 4). Despite the large grains, which display a rather constant Cu-Co composition of 20-30 at. % Co, regions with a composition of considerably higher and lower Co or Cu concentration are visible. These substructures have been also observed in the BSE micrographs. In the high-spatial resolution EELS Cu and Co mappings displayed in Fig. 8c-d, nc grains with constant gray value are visible. At their grain boundaries, alternating Cu and Co rich particle and areas with a dimension of about 10- 50 nm can be seen. The average foil-thickness is assumed to be in a range of 20-50 nm, making it impossible to measure the composition of the Co- and Cu-rich regions accurately, as EELS maps provide only a 2D-projection of the microstructure. To perform precise composition measurements and to gain more information about the 3D-morpology and the composition of the small scaled sub-structured regions, APT analyses were performed, additionally. As expected, the elemental distribution is inhomogeneous in the annealed state and the same distinct regions observed by analytical TEM can be identified in the APT reconstructions as seen in Fig. 9a (the presented elemental maps from the annealed condition are from two different APT specimens). Areas, in which regions with higher and lower Cu and Co concentration vary on a very fine scale are visible in both elemental maps. Between them, distinct Co and Cu enriched regions can be observed. From these regions local compositional measurements were carried out (indicated by the letters A-F), which showed that the Cu rich regions contain more than 99 at.% Cu, while the composition of the Co rich regions is above 96 at.%. The 1-dimensional (1D)-concentration profiles shown in Fig. 9b are obtained within a volume of 2 x 2 x 25 nm³ within the middle of the reconstructed sub-volumes "as-deformed" and "annealed" (obtained from Cu-Co alloy samples in the as-deformed and annealed state, respectively). The 1D-concentration profile of the as-deformed state exhibits only small local fluctuations of the Cu and Co concentration around their overall values of 71.5±5.6 at.% and 26.6±5.4 at.%, respectively. However, the annealed state indicates distinct variations in the Cu and Co concentration. Cu-rich regions with concentrations about 85 at.% become apparent. In between, Co- and Cu-rich regions with average concentrations of about 60 at.% are visible.

In Fig. 9c, the cumulative increase in the number of Co and Cu atoms with respect to the total number of atoms from the same regions of interest as shown in Fig. 9b are plotted. In a cumulative plot a constant slope corresponds to a respective phase in the material. Hence, the slope m of the corresponding curve is equivalent to the local concentration and an increase in slope indicates regions of increased Cu or Co concentration, respectively. In the as-deformed state, a constant slope is determined for Cu and Co in the cumulative distribution plot. In contrast, different slopes revealing regions of significantly increased and decreased Cu and Co concentrations are visible in the cumulative distribution plot in the annealed state.

To assess morphology, connectivity and sharpness of the modulated regions, isoconcentration surfaces are constructed within a representative volume of 17 x 28 x 35 nm³ for the overall composition of the alloy and at higher concentration values, respectively. Therefore, the isoconcentration surfaces displayed in Fig. 10a are drawn at elemental concentration values of 0 at.%, 5at.%, 10 at.% and 20 at.% above the overall composition of the Cu-Co alloy. The Cu-and Co rich areas construct a 3D interconnected network, in which the different areas alter within a few nanometers. The decrease in volume fraction and size of the corresponding Cu- and Co-enriched domains with increasing threshold values suggest that the Cu- and Co- rich areas are separated by diffuse boundaries. For threshold values of 75 at.% Co, no isoconcentration surfaces can be created, suggesting that no detectable Co precipitates are present in these areas. The elemental map in Fig. 10b show Co enriched regions with a size of less than 10 nm located next to a fine-compositional modulated region. 1D-concentration profiles are obtained from two regions marked A and B as indicated in the elemental map. The Co- and Cu-rich regions consist of almost pure Co and Cu in both concentration profiles, which is close to the equilibrium composition. In the modulated regions, alternating Cu and Co rich domains are visible in which the average composition of Co does not exceed 60 at.%. These findings can be again visualized by corresponding cumulative Co and Cu distribution plots of the same respective data, in which the total number of detected ions (Cu + Co) is plotted versus the number of detected Co and Cu ions, respectively. Through all Co and Cu particles, a nearly constant slope m is determined.

## 4. Discussion

The results of the structural investigations show that isothermal annealing leads to the development of a complex microstructure in the Cu-Co alloy. Besides some grains without considerable contrast variations, a morphology of a 3D network of alternating Cu and Co rich areas develops in the vast majority of the grain interiors after annealing at intermediate annealing temperatures. Pure Cu and Co particles are additionally formed near the grain boundary regions as revealed by APT and TEM, which are almost exclusively arrayed in alternate arrangement as it is clearly evident from the TEM micrographs in Fig. 8. Their length scale covers small, ~10 nm, to larger particles, ~50 nm, of nearly pure Cu and Co.

From DSC analysis and microstructural investigations as function of annealing temperature and time of the Cu-Co alloy, it is evident that at low annealing temperatures ($\leq 250°C$) only recovery takes place. No significant grain growth occurs even for long annealing times. At comparable annealing temperatures, pronounced grain growth is found for the pure Cu reference material even after a short annealing time of 1h. Grain growth in pure Co occurs at slightly higher temperatures. This shift in grain growth temperature can be explained by the higher melting temperatures of Co compared to Cu. The activation energies for grain growth $Q$ from Kissinger analysis for pure Cu and Co, which are 0.96±0.05 eV and 1.83±0.12 eV, further reflect this difference. The value found for Cu is nearly identical to those found for grain growth in HPT deformed Cu reported by different research groups [50-52]. It is furthermore very close to the activation energy for grain growth of nc Cu prepared by different methods [53,54]. The increase in grain growth temperature of the Cu-Co solid solution compared to pure Cu indicates that Co additions enhance the stability of the Cu phase. As long as Co atoms are dissolved in the *fcc* Cu-Co solid solution lattice, these solutes might hinder the migration of *fcc* Cu grain boundaries due to solute drag. Hence, no pronounced grain growth can be observed before the Co atoms are removed from these solid solution.

Upon annealing at higher temperatures, decomposition of the Cu-Co solid solution starts. From the DSC scans, a massive heat release in a temperature range between 250°C and 600°C is observed. Temperatures previously reported for the onset of phase separation in Cu-Co alloys prepared by rapid quenching and mechanical alloying are also located in these temperature range [29,33]. Pronounced grain growth in the Cu-Co alloy is only observed at temperatures higher than 600°C in the non-isothermal annealed samples (Fig. 1c). Nevertheless, grain growth occurs even before the onset of the second, large exothermic peak observed in the DTA scan. The coarsening of the microstructure during isothermal annealing with increasing time at 600°C is clearly evident in the BSE and TEM micrographs (Fig. 4, Fig. 6), which is a further indication that decomposition of the Cu-Co solid solution takes already place between 250°C and 600°C. For the Cu-Co alloy samples, the peak temperatures $T$ of the main exothermic peak in the DSC scans at various heating rates are also evaluated by Kissinger analysis (Table 1). The nature of the subsequent, weak exothermic DSC peak is not yet clear. Due to the large width of the poorly separated peaks, the kinetics of decomposition further cannot be modeled by a single and simple nucleation and growth process. Hence, the validity of the Kissinger analysis in terms of a single process is not strictly fulfilled and the activation energy, 1.79±0.03 eV, can thus only give a qualitative hint for the underlying processes occurring during the exothermic event. The literature data for different activation energies for Co and Cu diffusion are given in Table 2. The value of activation energy obtained from the present results is lower than values typical found for Co tracer diffusion in bulk Cu, 2.2 eV [55], and Co-Cu interdiffusion coefficient, 2.5 eV [56]. However, the measured activation energies for the Cu-Co alloy (1.79±0.03 eV) and the pure Co reference material (1.83±0.12 eV) are nearly identical.

It is believed that the decomposition process in the HPT deformed Cu-Co alloy does not occur via a classical nucleation and growth mechanism. The overall composition of the Cu-Co alloy (26 at.% Co) is well within the spinodal (~5 at.% Co at 400°C) [28]. As a consequence, the formed solid solutions are instantly unstable after HPT deformation. Although Co shows an extended solubility in Cu after HPT deformation, these solid solutions are not homogenous at the atomic level and Co is enriched in nanometer-sized clusters in the fcc Cu phase even in the as-deformed state (for further details see Ref. [11]). This is not unexpected since the total free energy decrease immediately after HPT deformation if Co- and Cu-rich regions with small fluctuations in composition are formed. From the development of the isoconcentration surfaces with increasing threshold values (Fig. 10a) and the 1D-concentration profiles of the as-deformed and annealed state, diffusion against the concentration gradient is suggested and it seems that decomposition proceeds similar to spinodal decomposition. The wavelength and amplitude are common parameters to describe spinodally decomposed alloys [61,62]. For the Cu-Co system, a decomposition on a very small length scale of less than 2 nm during annealing at 450°C is theoretically predicted [63,64]. Although the 1D-concentration profiles from the modulated regions can, in principle, provide the wavelength and amplitude of spinodal decomposition, it is difficult to align the 1D profile along the spinodal propagation. In order to investigate the underlying mechanisms of phase separation more closely and prove spinodal decomposition, it would be of course necessary to evaluate also the early stages of decomposition in this alloy system.

Although there exist no thermodynamic barrier for decomposition of an alloy with a composition inside the spinodal, the kinetics of the process depend strongly on annealing temperature [61]. The mobility of the solute Co atoms in the Cu matrix, which are responsible for the observed strong temperature-dependent decomposition kinetics, is drastically reduced at lower annealing temperatures. A difficulty is that there is limited amount of information available in the literature on diffusion of Co in Cu at low temperatures. The diffusion distances $L$ as function of annealing time $t$ can be roughly estimated as $L \approx (Dt)^{1/2}$ and are presented in Table 3 using the interdiffusion coefficient of Cu-Co and Co tracer diffusion coefficient in Cu at a temperature of 300, 400 and 600°C by extrapolating the Arrhenius equation to lower temperatures [55,56]. In the maximum given time of 100h, cobalt diffuses a distance less than 1 nm at 300°C. Based on these estimates it can be concluded that practically no diffusion will occur within the grains at low temperatures, even for long time periods. As a consequence, nearly no compositional modulated sub-structures can be observed in the Cu-Co alloy at low annealing temperature ($\leq$ 300°C) as the atomic mobility of the Co atoms is too low. A temperature above 400°C is needed for any significant lattice diffusion. The time provided at the isothermal annealing temperatures of 400°C should kinetically allow the observed decomposition processes to take place. This is further supported by calculations, in which the time, which is needed for complete spinodal decomposition of a Cu-Co solid solution, is about 170 min at an annealing temperature of 450°C [63]. At higher annealing temperatures, the kinetic (diffusion) limitation of the decomposition processes is eliminated. A very short time of only less than 2 min is calculated to be necessary to reach Co equilibrium concentrations within the modulations at annealing temperatures of 600°C [63]. After phase separation occurred, the grain size on annealing at 600°C coarsens rapidly in the alloy. On the other hand, despite the observed grain growth, small Co particles can be still found at triple points and between larger grains (Fig. 6). These small particles act as additional dragging force for grain boundary migration, thus stabilizing the ufg structure against further massive grain growth at 600°C.

Pure, alternating Cu and Co particles, but with a significantly enhanced length scale compared to the grain interior, are additionally found near the grain boundary regions in the microstructure annealed at intermediate annealing temperatures. Two complementary processes might be considered for their formation: Pure Co precipitates can nucleate heterogeneously at the grain boundaries and subsequently grow in size. Alternatively, the same spinodal decomposition process, but with an enhanced wavelength, takes place near the grain boundary. There is strong evidence from APT data, that heterogeneous precipitation can be excluded in the studied Cu-Co alloy system. A series of cross-section elemental maps through the 3D reconstructed volumes of the APT specimen shown in Fig. 9a illustrate that the observed Co and Cu particles are located not directly at the grain boundary, but inside the corresponding grain separated by a thin Cu layer from the grain boundary (see Supplementary Fig. S2).

In HPT deformed materials, a large amount of vacancies together with a high dislocation density and highly-distorted non-equilibrium grain boundaries are created. In addition, recent studies showed that diffusion or segregation processes in ufg materials are considerably influenced by the highly-distorted non-equilibrium grain boundaries [65-68]. The vacancy-type free volumes are supposed to be mainly located in the grain boundary regions, thus enhancing grain boundary diffusion. After spinodal decomposition, coarsening by Ostwald ripening, which can occur in any two phase mixture with significant diffusion, might occur more easily in the grain boundary regions.

A characteristic of spinodal decomposition is that a number of concentration fluctuations with a wavelength larger than a critical wavelength $\lambda_c$ will develop on annealing. The fastest growing wavelength will be determined by $\lambda_{max} = \sqrt{2}\lambda_c$, whereas $\lambda_c$ is given by [69]

$$\lambda_c = 2\pi \left( \sqrt{-\left[\frac{\partial^2 G}{\partial c^2} + 2\delta^2 \left(\frac{E}{1-v}\right)\right] \frac{1}{2K}} \right)^{-1}, \qquad (3)$$

with $K$ being the gradient energy coefficient, δ the lattice misfit of CuCo, $\frac{\partial^2 G}{\partial c^2}$ the second derivate of the free energy with respect to the atomic fraction $x$ at a given temperature, $E$ the elastic modulus and ν the Poisson's ratio. The wavelength increasingly deviates for smaller compositions and becomes largest very close to the spinodes ($\frac{\partial^2 G}{\partial c^2} = 0$). Therefore, it might be also assumed that Co concentration variations differ from grain to grain (or even inside a grain), which can lead to the observed differences in the length of the compositional modulations. The analysis of different solubility levels from grain to grain in the as-deformed state is a topic worthy of further study and will be focus of some of our future work.

In [25] it is proposed that binary alloy combinations with a large segregation enthalpy relative to the enthalpy of mixing are generally the systems, which are expected to exhibit a stable nc state. Although the Cu-Co system has a positive enthalpy of mixing, calculations of the dilute-limit grain boundary segregation enthalpy for Co in Cu indicate an anti-segregation tendency (negative value of the segregation enthalpy). As a consequence, no stable nc state is expected in the alloy system studied in this paper by thermodynamic considerations [25-27]. Therefore, an ideal nanostructure as recently reported for a W-20at.% Ti alloy is not attained during annealing in our alloy system [70]. Nonetheless, it is shown in this study that the phase decomposition of the metastable Cu-Co solid solution has a significant influence on the thermal stability, which can be linked to the underlying microstructure that forms during annealing. In principle, such phase decomposition mechanisms can be used to produce tailored nano structures to optimize e.g. thermal or mechanical properties despite a stabilization by thermodynamic considerations.

## 5. Conclusion

The thermal stability of supersaturated Cu-Co solid solutions prepared by HPT is investigated and linked to the underlying microstructure that forms during annealing. Initial phase separation is a necessary prerequisite for grain growth. At elevated temperatures, structural relaxation takes place, but no grain growth prior to phase separation of the metastable supersaturated solid solution is observed even during long-term isothermal annealing up to 100 h. Annealing at 400°C results in a complex microstructure consisting of compositional modulated grains next to pure Co and Cu particles located near the grain boundaries and at triple points of the nc structure. The decomposition process inside the grains proceeds against the concentration gradient and hence indicate spinodal decomposition. Once the microstructure is formed and phase separation is nearly finished, it is stable even for very long annealing times of several days and exhibit both, high stability and high hardness at the intermediate annealing temperature. It is proposed that the Co and Cu particles near the grain boundary act as additional dragging force for grain boundary migration. In contrast, isothermal annealing at higher temperatures results in accelerated phase separation processes due to enhanced kinetics. Once the elemental Cu and Co phases are formed, grain growth occurs even at quite short annealing times and an ufg structure is formed.

**Acknowledgments**

The authors gratefully acknowledge the financial support by the Austrian Science Fund (FWF): J3468-N20. The atom probe instrument was financed by the DFG (INST 256/298-1 FUGG) and the Federal State Government of Saarland. We thank N. de Jonge and E. Arzt for support through INM. We further thank the Erich Schmid Institute of Materials Science in Austria for providing all HPT deformed samples.

**Tables**

Table 1: The peak temperatures of the exothermic peaks measured under varying DSC heating rates and activation energies for the pure Cu and Co reference and Cu-Co alloy samples.

|  | Heating rate Φ (°Cmin$^{-1}$) | Peak temperature T (°C) | Activation energy (eV) |
|---|---|---|---|
| Cu | 10 | 319 | 0.96±0.05 |
|  | 20 | 336 |  |
|  | 60 | 376 |  |
|  | 120 | 396 |  |
| Co | 10 | 456 | 1.83±0.12 |
|  | 20 | 477 |  |
|  | 60 | 504 |  |
|  | 120 | 519 |  |
| Cu-Co | 10 | 416 | 1.79±0.03 |
|  | 20 | 432 |  |
|  | 60 | 456 |  |
|  | 120 | 474 |  |

Table 2: Activation energies for different mechanisms of Co and Cu diffusion. The references from which the data is taken is indicated in the table.

| Mechanism | Activation energy (eV) |
|---|---|
| Cu self-diffusion [57] | 2.1 |
| Cu grain-boundary diffusion [58] | 0.8-0.9* |
| Co self-diffusion (fcc) [59] | 3.0 |
| Co grain-boundary diffusion [60] | 1.7 |
| Co tracer diffusion in Cu [55] | 2.2 |
| Cu-Co interdiffusion [56] | 2.5 |

*depending on purity of Cu polycrystals

Table 3: Diffusion distance L at different annealing times t and annealing temperatures (300°C, 400°C, 600°C) using the interdiffusion coefficient of Cu-Co [56] and Co tracer diffusion coefficient in Cu [55], respectively.

|  | t (h) | L(nm), 300°C | L(nm), 400°C | L(nm), 600°C |
|---|---|---|---|---|
| Cu-Co interdiffusion | 1 | 0.0 | 0.5 | 77.9 |
|  | 7 | 0.0 | 1.4 | 206.3 |
|  | 100 | 0.1 | 5.4 | 779.9 |
| Co tracer diffusion | 1 | 0.1 | 2.3 | 173.7 |
|  | 7 | 0.2 | 6.0 | 459.5 |
|  | 100 | 0.8 | 22.8 | 1736.9 |

**Figures**

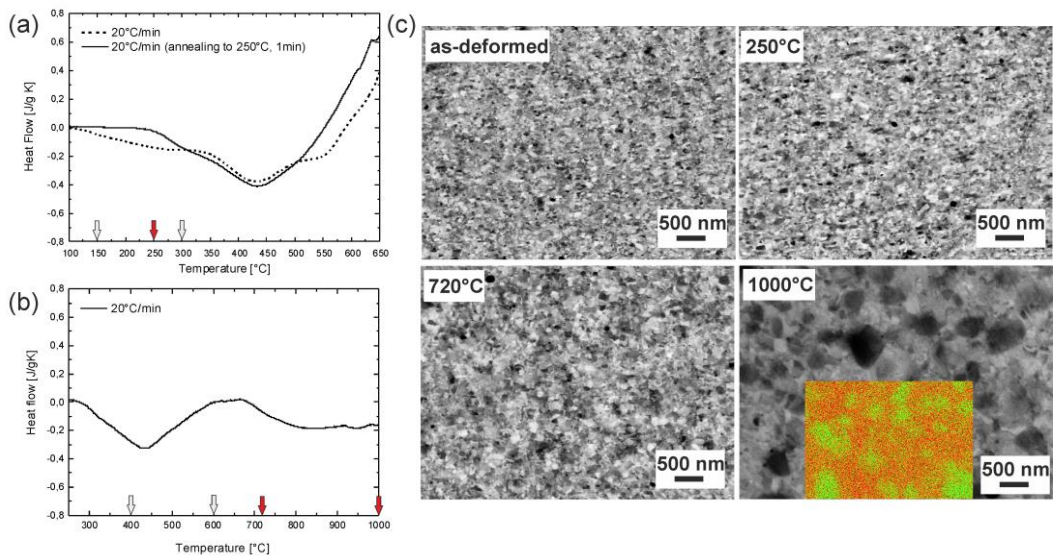

Fig.1: (a) Typical DSC scans for the Cu-Co alloy after HPT deformation (dotted line) and with a preannealing treatment (solid line) at a heating rate of 20°C/min. The preannealing was conducted by heating the as-deformed sample in the DSC at 20°C/min to a temperature of 250°C and rapidly cooling to room temperature. (b) DTA scan for the Cu-Co alloy after HPT deformation. Arrows in red and gray indicate annealing states for which SEM micrographs were taken and annealing temperatures for the isothermal long-term annealing investigations. (c) BSE micrographs of the microstructure of the as-deformed Cu-Co alloy sample and after annealing to predefined temperatures at a heating rate of 20°C/min: as-deformed, annealed to 250°C, 720°C and 1000°C. The inset shows a superimposed EDS map with all regions belonging to Co are marked as green and all regions belonging to Cu are red.

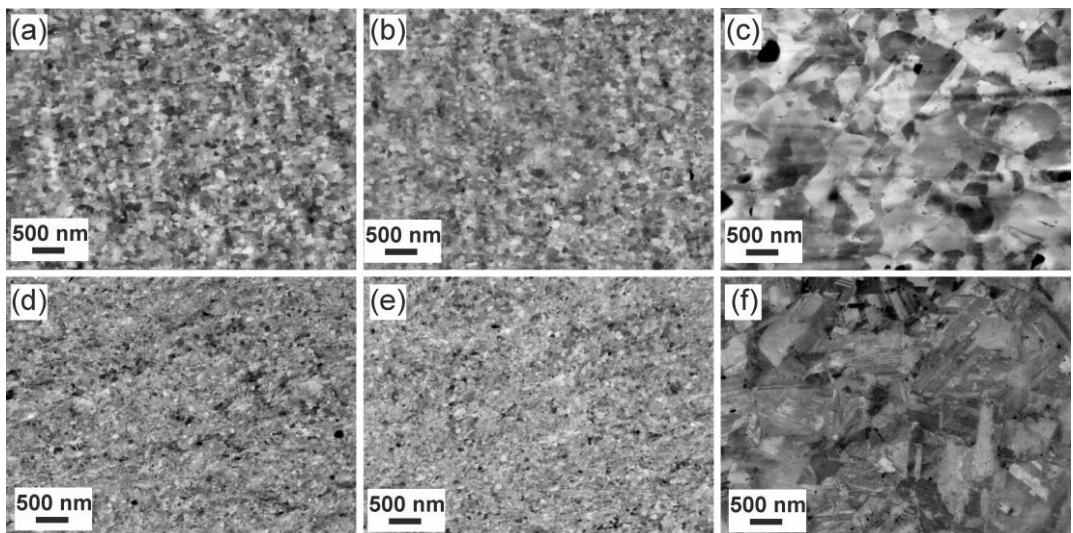

Fig. 2: BSE micrographs of the microstructure of the as-deformed Cu and Co samples and after annealing to predefined temperatures at a heating rate of 20°C/min: (a) as-deformed Cu sample, (b) Cu sample annealed to 250°C and (c) 720°C, (d) as-deformed Co sample, (e) Co sample annealed to 250°C and (f) 720°C. The magnification is the same in all micrographs.

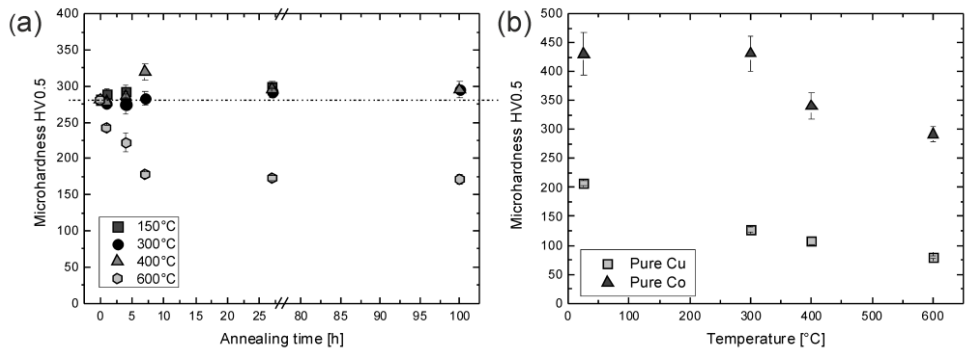

Fig. 3: (a) Microhardness (HV0.5) of the Cu-Co alloy as function of annealing time (0, 1, 4, 7, 27, 100 h) for different annealing temperatures (150°C, 300°C, 400°C, 600°C). (b) Microhardness (HV0.5) of pure Cu and Co as function of annealing temperature (300°C, 400°C, 600°C) after 1h annealing time.

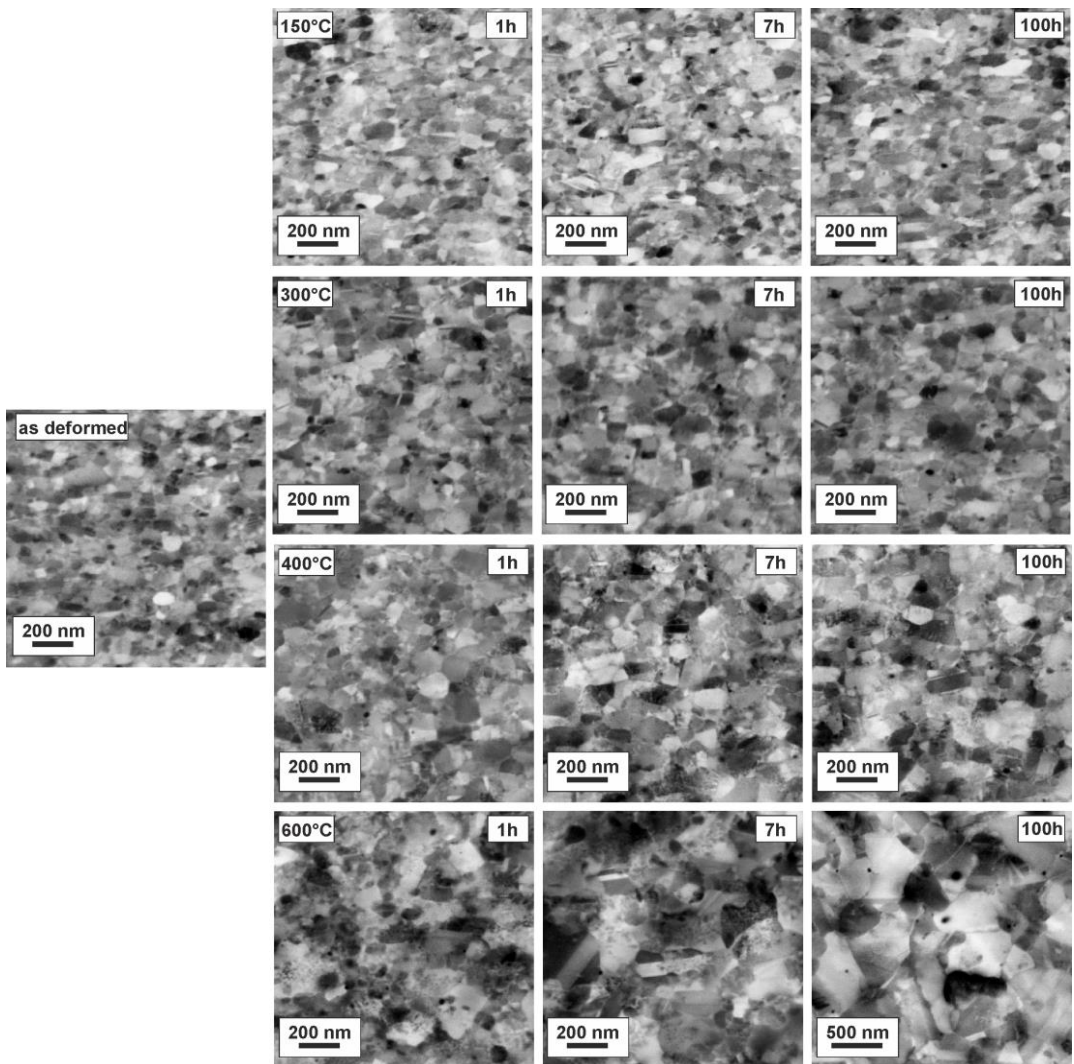

Fig. 4: BSE micrographs of the microstructure of the Cu-Co alloy after HPT deformation and annealing for 1h, 7h and 100h at 150°C, 400°C and 600°C. The magnification is the same for all micrographs except for the sample annealed at 600°C for 100h.

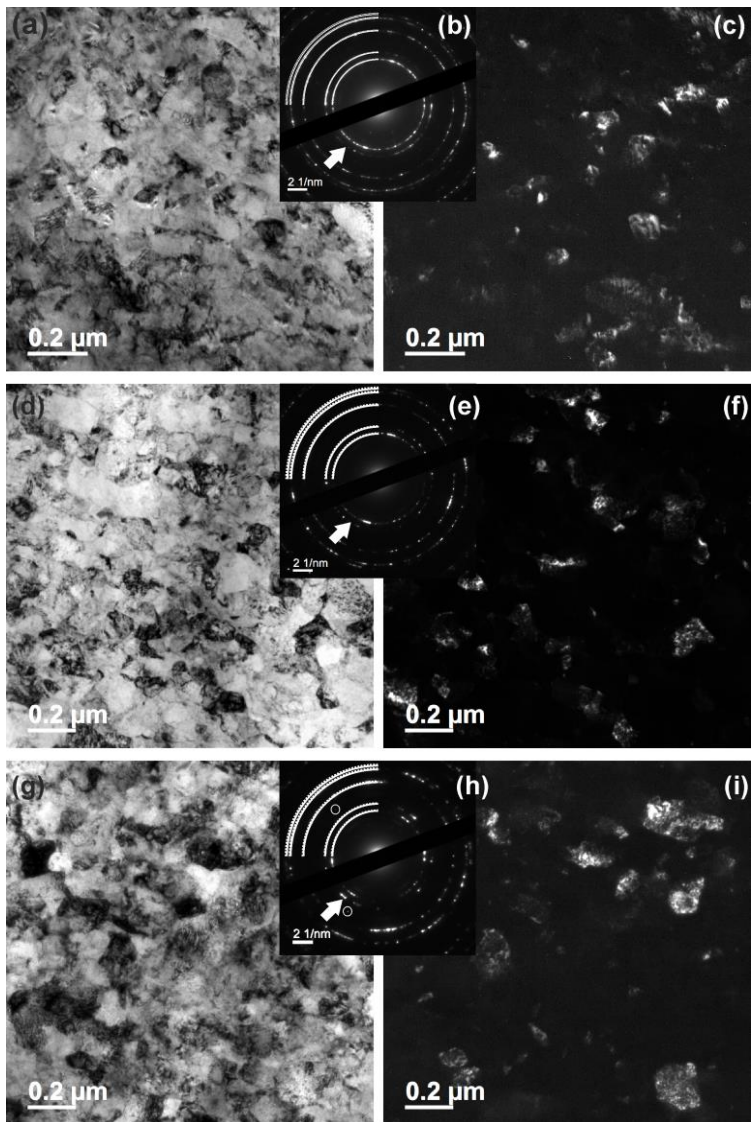

Fig. 5: (a) BF TEM micrographs of the Cu-Co alloy after HPT deformation and annealing for 100h at (d) 300°C and (g) 400°C. SAD patterns with reflexes of *fcc* Cu (solid lines) and fcc Co (dotted lines) are shown for the (b) as-deformed and annealed samples ((e) 300°C and (h) 400°C). Cu reflections (111), (200), (220), (311), (222); *fcc* Co reflections (111), (200), (220), (311), (222). DF micrographs of the (c) as-deformed and annealed samples ((f) 300°C and (i) 400°C) of the same area obtained by selecting some (1 1 1) Cu or (1 1 1) Co lattice reflections (first Debye–Scherrer ring).

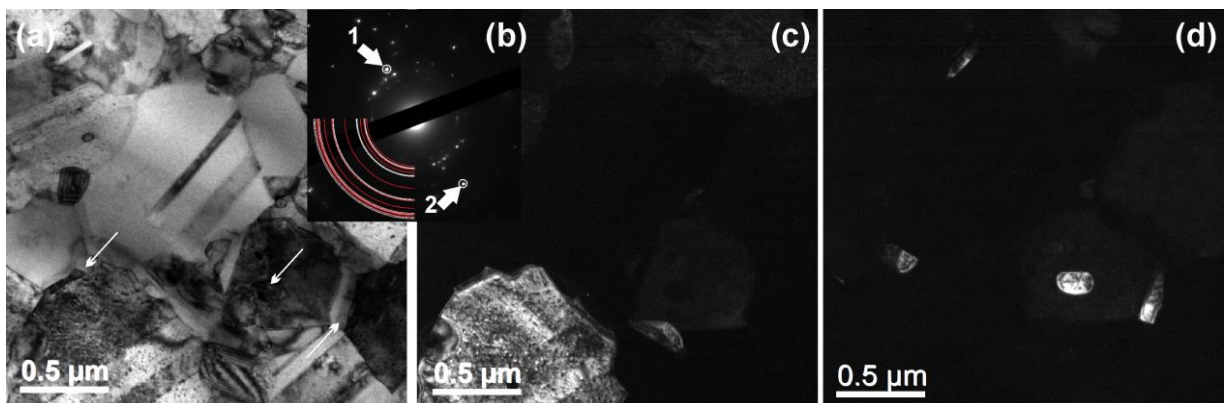

Fig. 6: (a) BF micrographs of the Cu-Co alloy after annealing for 100h at 600°C. (b) SAD pattern with reflexes of *fcc* Cu (solid lines), *fcc* Co (dotted lines) and *hcp* Co (dashed lines). Cu reflections (111), (200), (220), (311), (222); *fcc* Co reflections (111), (200), (220), (311), (222); *hcp* Co reflections (100), (002), (101), (102), (110), (103), (112), (201). (c) DF micrograph of the same area obtained by selecting some (1 1 1) Cu or (1 1 1) Co lattice reflections (indicated as "1" in the SAD pattern) and (d) by selecting (220) *fcc* Co or (110) *hcp* Co lattice reflections (indicated as "2").

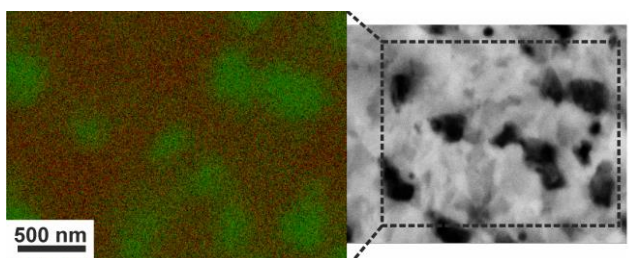

Fig. 7: BSE micrograph of the Cu-Co alloy sample after annealing for 100h at 600°C. In the EDS map, all regions belonging to Cu are marked as red and all regions belonging to Co are marked as green.

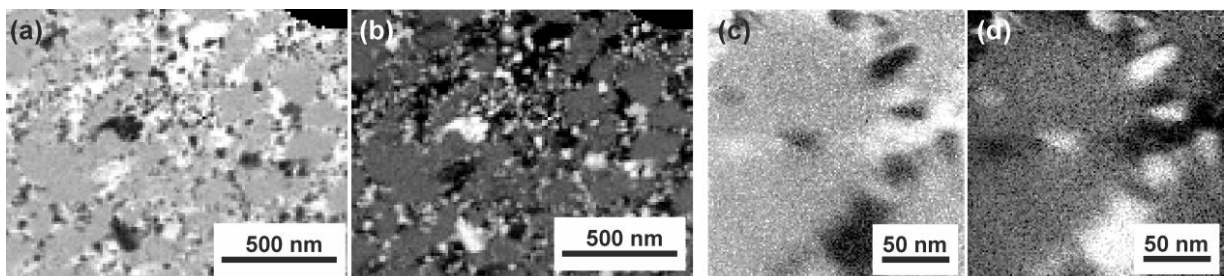

Fig. 8: EELS Cu (a) and Co (b) maps (size 1500 x 1200 nm²) of the Cu-Co alloy sample after annealing for 100h at 400°C. High-spatial resolution EELS Cu (c) and Co (d) maps (200 x 215 nm²).

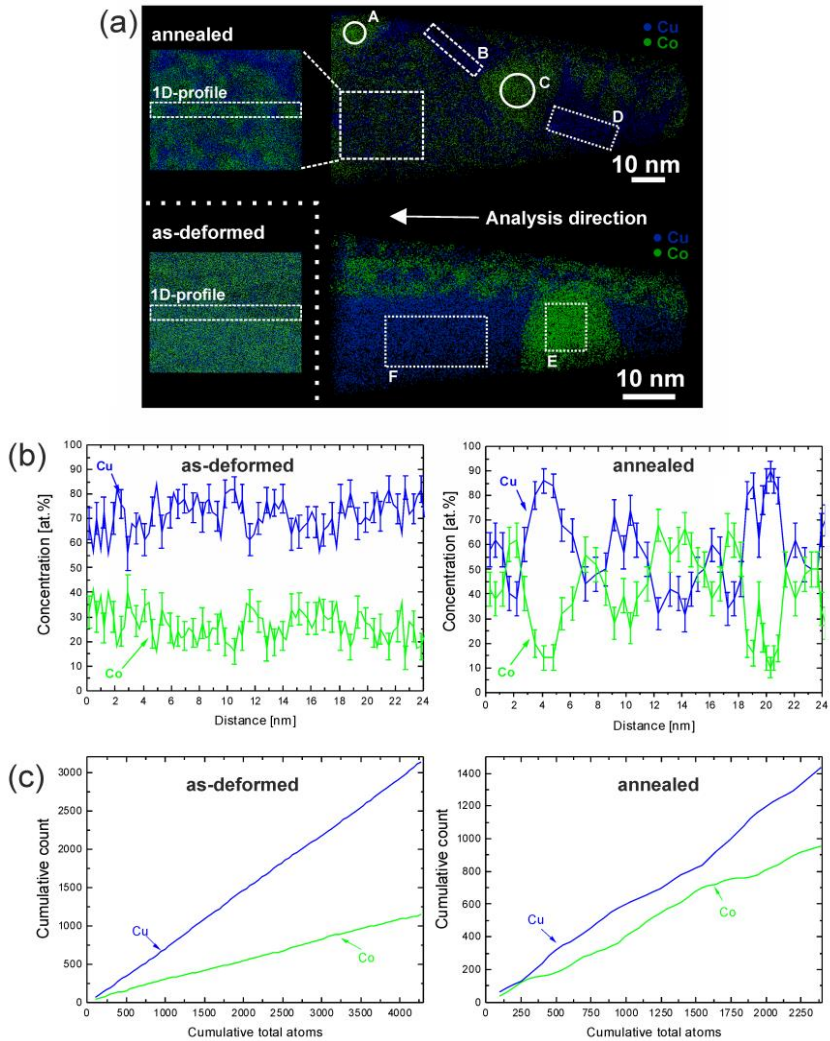

Fig. 9: (a) APT elemental maps from three different reconstructions. The two reconstructions on the right are from two APT specimens of the Cu-Co alloy sample annealed at 400°C for 100h. The maps shown here are 10 nm thickness slices (perpendicular to the figure) from the complete reconstructions. Different regions-of-interest (A-F) are selected for composition measurements which are marked in the reconstructions. Size of sampling spheres: Ø 8.8 nm (A), Ø 10 nm (C). Size of sampling cylinders: Ø 4 nm x 20 nm (B), Ø 7.5 nm x 20 nm (D). Size of sampling cubes: 7.5 x 12 x 9 nm³ (E), 10 x 10 x 20 nm³ (F). The third reconstruction is from an as-deformed specimen where only a sub-volume (10 x 20 x 25 nm³) is shown here (bottom left). A same volume is extracted from the annealed specimen dataset (top left) for profile concentration comparison. (b) 1D-concentration profiles computed along the region-of-interest marked as "1D-profile" in the reconstructed sub-volumes "as-deformed" and "annealed". (c) Cumulative Co and Cu distribution plots computed along the same regions of interest as marked in (b) in the reconstructed sub-volumes "as-deformed" and "annealed" (For interpretation of the references to color in this figure legend, the reader is referred to the web version of this article).

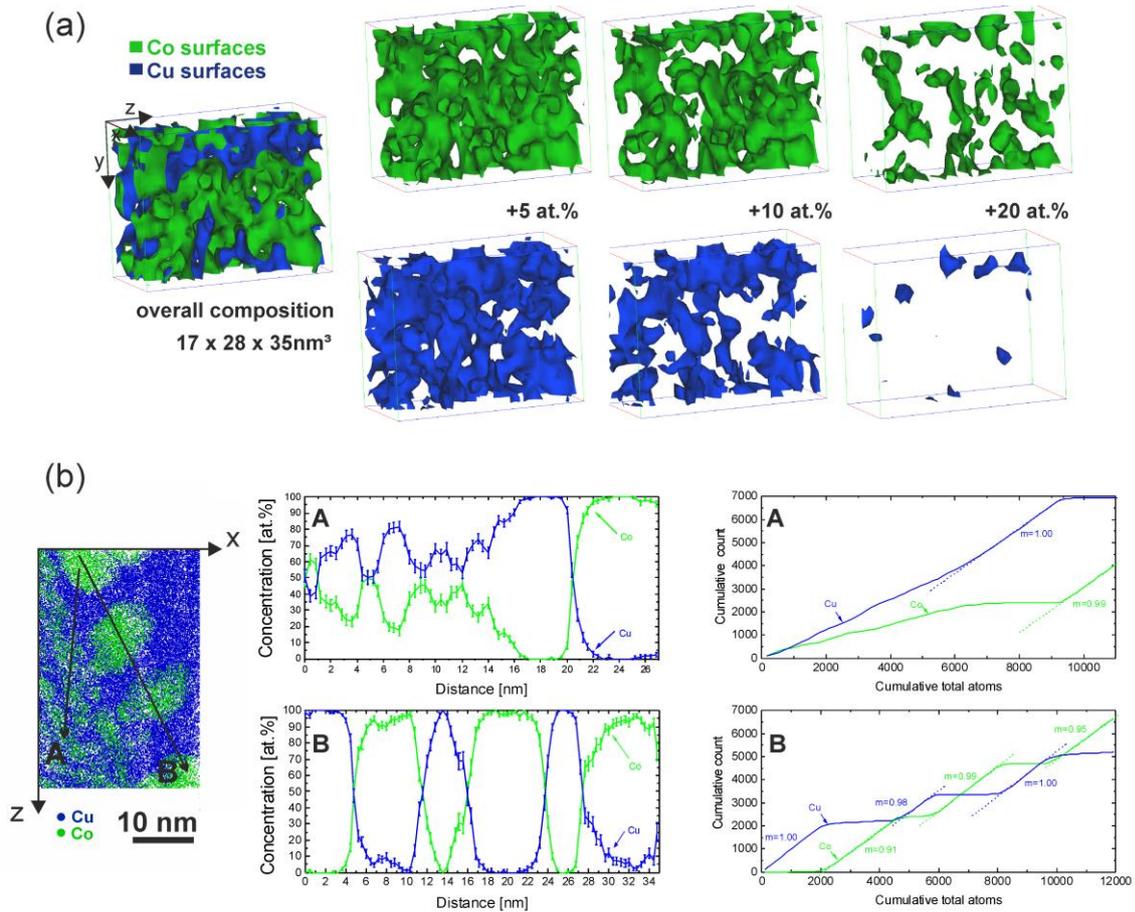

Fig. 10: (a) Combined isoconcentration surfaces of Co (green) and Cu (blue) within a selected volume of 17 x 28 x 35 nm³ at threshold values for the overall compositions of Co (28 at.%) and Cu (72 at.%). Isoconcentration surfaces of Co and Cu with increasing threshold values by 5 at.% (Co-33at.%, Cu-77at.%), by 10 at.% (Co-38at.%, Cu-82at.%) and by 20 at.% (Co-48at.%, Cu-92at.%). (b) Elemental distributions of Cu and Co of the Cu-Co alloy sample (annealing temperature of 400°C for 100h). The thickness of the reconstructed volume (perpendicular to the figure) is 8 nm. 1D-concentration profiles and corresponding cumulative Co and Cu distribution plots computed along the lines marked as A and B in the reconstructed volume (For interpretation of the references to color in this figure legend, the reader is referred to the web version of this article).

**Supplementary Figures**

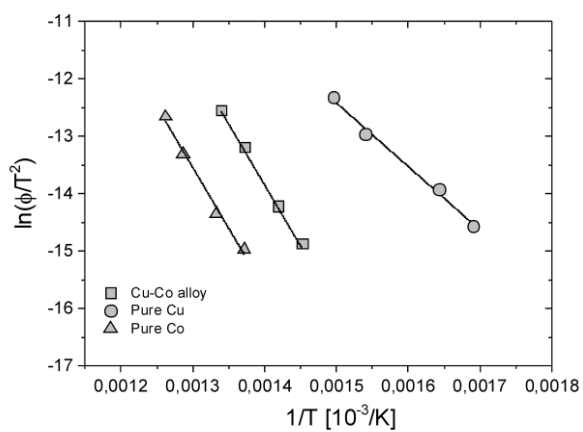

Sup. Fig. S1: Kissinger plot of the temperatures *T* measured on pure Cu, pure Co and the Cu-Co alloy samples applying different heating rates Φ in the range from 10 to 120 °C/min.

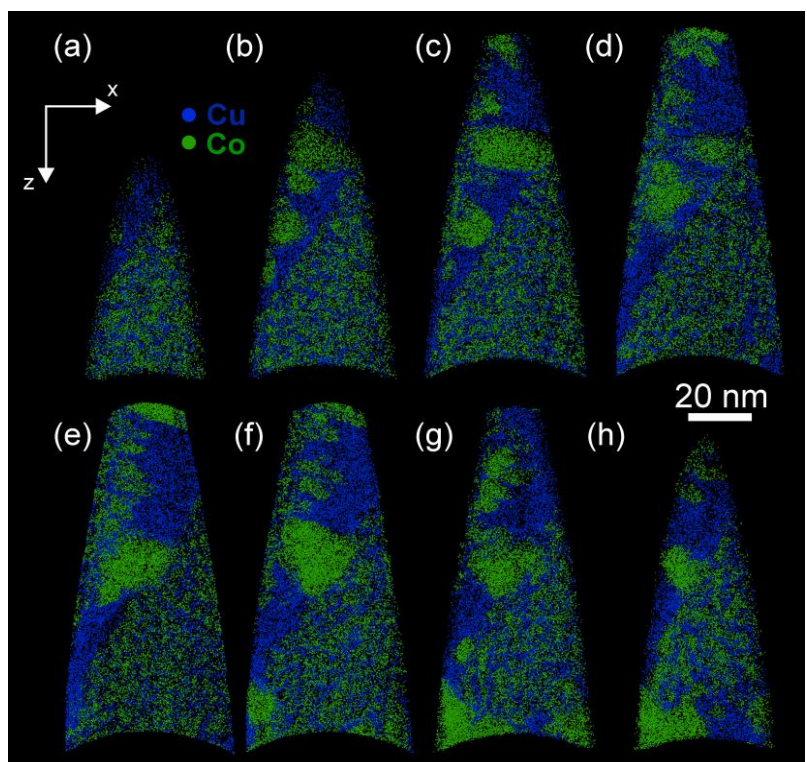

Sup. Fig. S2: 2D cross-section elemental maps ((a)-(h)) through the 3D reconstructed volumes of the APT specimen of the Cu-Co alloy sample annealed at 400°C for 100h. The thickness perpendicular to the figure is 5nm in each cross-section, respectively (For interpretation of the references to color in this figure legend, the reader is referred to the web version of this article).